\documentclass[preprints,hypothesis,submit,moreauthors]{Definitions/mdpi} 
\firstpage{1} 
\pubvolume{1}
\issuenum{1}
\articlenumber{0}
\pubyear{2023}
\copyrightyear{2023}
\datereceived{N/A} 
\dateaccepted{N/A} 
\datepublished{N/A} 
\hreflink{https://doi.org/to-be-added} 

\usepackage{subcaption}

\Title{Generative AI in Writing Research Papers: A New Type of Algorithmic Bias and Uncertainty in Scholarly Work}

\TitleCitation{Generative AI in Writing Research Papers: A New Type of Algorithmic Bias and Uncertainty in Scholarly Work}

\Author{Rishab Jain $^{1,2 }$\orcidA{}, Aditya Jain $^{3 }$\orcidB{}}

\AuthorNames{Rishab Jain, Aditya Jain}

\AuthorCitation{Jain, R., Jain, A.}

\address{$^{1}$ \quad Harvard University, Cambridge, MA 02138, USA States; \\
$^{2}$ \quad Department of Neurosurgery, Massachusetts General Hospital, Boston, MA 02114, USA; \\
$^{3}$ \quad Harvard Medical School, Boston, MA 02115, USA}

\corres{Correspondence: rishabjain@college.harvard.edu}

\abstract{The use of artificial intelligence (AI) in research across all disciplines is becoming ubiquitous. However, this ubiquity is largely driven by hyperspecific AI models developed during scientific studies for accomplishing a well-defined, data-dense task. These AI models introduce apparent, human-recognizable biases because they are trained with finite, specific data sets and parameters. However, the efficacy of using large language models (LLMs)---and LLM-powered generative AI tools, such as ChatGPT---to assist the research process is currently indeterminate. These generative AI tools, trained on general and imperceptibly large datasets along with human feedback, present challenges in identifying and addressing biases. Furthermore, these models are susceptible to goal misgeneralization, hallucinations, and adversarial attacks such as red teaming prompts---which can be unintentionally performed by human researchers, resulting in harmful outputs. These outputs are reinforced in research---where an increasing number of individuals have begun to use generative AI to compose manuscripts. Efforts into AI interpretability lag behind development, and the implicit variations that occur when prompting and providing context to a chatbot introduce uncertainty and irreproducibility. We thereby find that incorporating generative AI in the process of writing research manuscripts introduces a new type of context-induced algorithmic bias and has unintended side effects that are largely detrimental to academia, knowledge production, and communicating research.}


\begin{document}

\section{Introduction}
Research that utilizes artificial intelligence (AI), machine learning, or deep learning technologies have shown an exponential growth in terms of the gross number of publications and citations \cite{aireviewpaper, Frank2019}. However, not all aspects of AI research are created equally: some fields employ AI more than others, and for different purposes. Medicine, for instance, has utilized AI as a tool for predicting clinical outcomes, whilst philosophy has used AI as a thought experiment to explore arguments about morality \cite{Frank2019}. Generative AI---now capable of writing human-like text---has been most recently shaped by the advent of large language models (LLMs) such as GPT-3 \cite{gpt3}. LLMs are trained artificial neural networks that can generate large outputs based on seed text. They can predict the next token in sequence based on the tokens before it, and are able to effectively generate text that is statistically similar to the human language it is trained on. Because LLM chatbots are trained on large swaths of human-written natural language (i.e. data from OpenWebText, Wikipedia, and online journalism), they incorporate human biases in their outputs \cite{ray_chatgpt_2023}.

While the biases present for AI models created for a certain domain---e.g. machine learning on the electronic health record \cite{gianfrancesco_potential_2018}---may be more easy to intuit, generative AI can be more unpredictable \cite{peng2023study, gloria2022bias}. Generative AI may have biases lurking under the surface, making it difficult to reproduce and definitively diagnose bias \cite{scheurer2023technical,mehrabi_survey_2022,noauthor_dangers_2021}. These may be data-driven or purely algorithmic, caused by training methods and model architectures \cite{noauthor_dangers_2021}. These biases are more concerning when factoring in hallucinations and goal misgeneralization: some of the biggest problems in AI interpretability and safety research \cite{scheurer2023technical,shah_goal_2022}. Generative AI is subject to hallucinations in which it may present information upon request with certainty, meanwhile it is wholly incorrect or nonsensical \cite{hallucination1,bang2023multitask,oniani_military_2023}. This can mislead humans into believing an algorithm has considerable understanding about the hallucination subject. Hallucinations become more likely if generative AI is trained on data subject to aggregation or representation bias \cite{dziri_origin_2022,mehrabi_survey_2022}. Goal misgeneralization may result in similar outcomes---particularly when generative AI chatbots play along stereotypes and assumptions about a user, based on their prompt \cite{casper_open_2023, casper_explore_2023}. Interpretability of AI cannot currently account for the many nuances and decisions a human may take when constructing or formatting a research paper. 

The online release of ChatGPT---powered initially by GPT-3 \cite{gpt3}---allowed the public to obtain coherent, AI-generated text with few-shot prompting \cite{ray_chatgpt_2023}. This has prompted analyses into ChatGPT's ability for human writing tasks: from essays to scholarly writing \cite{basic_chatgpt-35_2023,kumar_analysis_2023,alkaissi_artificial_2023,article1,article2,article4,article5,article6,article7,article8}. Biases in scholarly writing are present regardless of whether or not AI is used, however, humans can attempt to mitigate and contextualize the extent to which they introduce bias in their research \cite{pannucci_identifying_2010}. As chatbot-produced text can be generated with little to no human input, AI has the potential to automate scholarship and push out humans’ organic interpretation and decision making processes \cite{desaire_chatgpt_2023,article5,gandhi_does_2023}. Specifically, the use of generative AI in writing manuscripts yields papers that may not be original, pose ethical dilemmas, and introduce new types of algorithmic bias and uncertainty \cite{article6,article7,gandhi_does_2023}. Due to the pressure put on scholars to publish, they may increasingly rely on automative solutions that are capable of producing fast yet potentially problematic papers \cite{miller_publish_2011}.

Recent AI safety research has showcased the dangers of red teaming and adversarial attacks against LLMs: humans prompt models with oftentimes innocuous text intending to disrupt the system’s algorithms \cite{perez_red_2022}. When conducting research, especially in providing large prompts with scientific data and text, a user may unknowingly generate a red teaming prompt by not appropriately providing context and having specificity. This may cause the deployed LLM to leak information about its own training data, or generate harmful, biased text \cite{perez_red_2022}. In regard to ChatGPT, models such as GPT-3 have been trained extensively and aligned to generate text from a neutral lens, however, this neutral lens can break via red teaming, resulting in biases, misconceptions, and harmful outputs \cite{casper_explore_2023}. Because of reinforcement learning from human feedback (RLHF) techniques, models can be rewarded via human feedback for reinforcing existing stereotypes and beliefs of misaligned humans \cite{casper_open_2023}. By reinforcing an individual's bias, a human oracle may be more likely to positively assess a model's performance---thus resulting in deceptive models that can amplify bias and mislead users into having satisfactory experiences, but not necessarily scientifically accurate ones. There are numerous challenges in collecting diverse human feedback that is inclusive and appropriate to combat the risk of societal biases being instilled via RLHF and goal misgeneralization \cite{noauthor_dangers_2021, casper_explore_2023, casper_open_2023}. These issues can also negatively affect researchers, where, depending on a researcher's provided prompt---which may include demographic information or ideological beliefs---text generated by a generative AI chatbot could discriminate against and lead researchers down different paths.

This paper finds that the use of generative AI to write research manuscripts introduces a new type of bias driven by user context, and reaffirms algorithmic bias that exists within large language models that is problematic to academia and science communication. We highlight how researchers from any demographic can unintentionally prompt a generative AI chatbot with personal information, allowing it to generate biased text in a way that contributes to the reinforcement of existing societal biases. Due to hallucinations and red teaming, generative AI chatbots can generate incorrect and harmful text, leading to problematic outcomes for researchers. Using generative AI as a tool for producing papers would thus require an effort in counter-bias, especially in authorship. We suggest that the wider research community explore the implications of a positive feedback loop in which generative AI is trained on partially machine-written papers, resulting in amplifications of data-driven and context-induced bias.

\section{Writing Papers with AI}
\begin{figure}[H]
\includegraphics[width=0.8\textwidth]{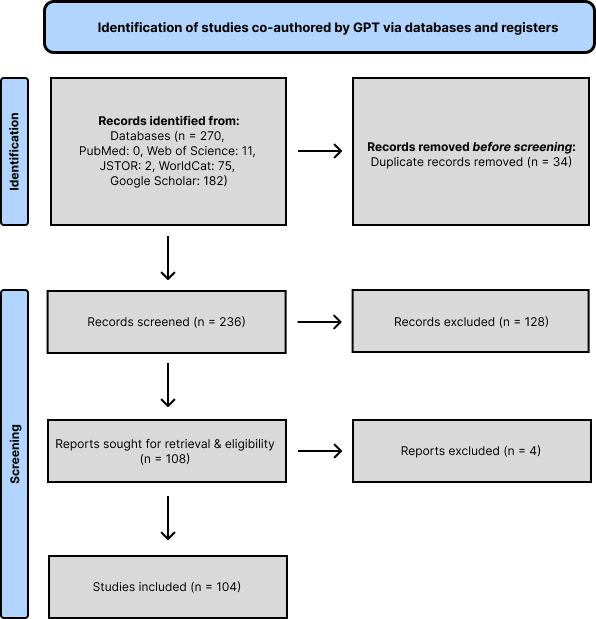}
\centering
\caption{We find 104 scholarly works using our criteria and query (Author: "ChatGPT" OR "GPT-2" OR "GPT-3" OR "Generative AI" OR "Generative Pretrained Transformer" OR "OpenAI").}
\label{fig:2}
\end{figure}

Generative AI technologies shaped by large language models, such as GPT-3, have changed the way humans interact with machines when communicating research. The most apparent use case is through few-shot prompting to write the contents of a research manuscript. This has been apparent with language models being added as co-authors on studies \cite{article8}. We quantify the number of manuscripts authored or co-authored by generative AI (more specifically, some variant of ChatGPT---the most well-known tool) by conducting a systematic review. We follow the Preferred Reporting Items for Systematic Reviews and Meta-Analyses (PRISMA) guidelines, using PubMed, Web of Science, JSTOR, WorldCat, and Google Scholar. We find 104 works as of November 2023 as shown in Figure \ref{fig:2} by querying the aforementioned databases and manually screening our results to ensure the identified pieces were authored by a generative AI tool.

\begin{figure}[H]
\includegraphics[width=0.68\textwidth]{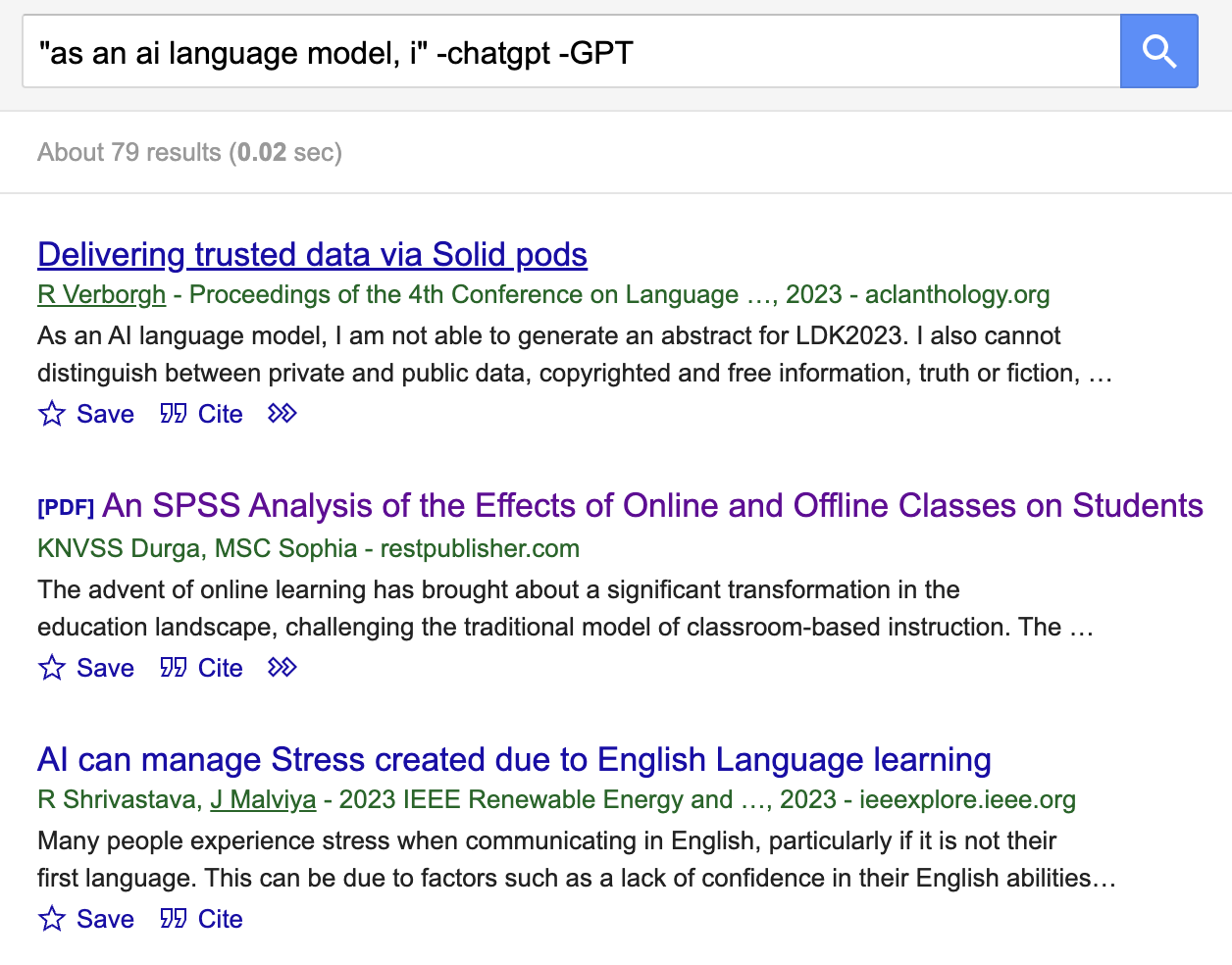}
\centering
\caption{We find 79 scholarly works that used ChatGPT and left its signature "As an AI language model" without edits.}
\label{fig:1}
\end{figure}

\noindent We note that remnants of AI prompts left in manuscripts due to careless errors indicate that our systematic review leaves out articles that used a generative AI tool, but neglected to cite it as an author. We verify this claim via Google Scholar's search engine, using the query as shown in Figure \ref{fig:1}. We suspect that far more authors have used ChatGPT similarly, but will have edited out such apparent markers of direct AI use---which some may deem plagiarism, depending on journal policy.

Journals and publishers have had widely different stances on authorship of AI-generated works \cite{article5}. In several open letters to journals, scientists have expressed their concerns about accepting AI-generated works \cite{gandhi_does_2023, article1}. These criticisms include whether AI-generated works are ethical, and whether a machine has the necessary intellectual ability to contribute originality and thereby be recognized as a legitimate author. While still in formation, several journals have centered their positions around authorship qualification: authors should only be added if they meet certain criteria \cite{article5,article1}. For instance, ChatGPT producing an abstract or summary when prompted with the rest of a paper does not make it an author, as authorship requires \textit{originality}. \textit{Science}'s Family of Journals prohibits the use of text generated by AI in their works, deeming it plagiarism \cite{article1, lin_supercharging_2023}. As of November 2023, both \textit{Science} and \textit{Nature} do not allow LLMs to be recognized as co-authors \cite{article1}. Largely, journals and concerned researchers have consensed that generative AI is not at the stage yet in which it can replace researcher originality. 

Researchers, however, may still find themselves increasing their dependence on AI due to other forms of AI assistance. AI can turn skeleton outlines into full papers, and also serve as auto-complete engines for more experienced authors \cite{lin_supercharging_2023}. Human-AI collaborative writing can be seen as an extension of the literature review process, in which humans enter keywords and AI generates text for the author to refine and add context \cite{aydin_openai_2022}. AI's knowledge of grammar rules, sentence structure, and word choices gives it a big advantage over human writing, especially for non-English speakers \cite{lin_supercharging_2023}. While this may help democratize access to publishing in English-based, high-impact journals, it also introduces a new set of problems. With humans growing accustomed to using AI for writing, the decisions and biases that AI produces become more entrenched in research. If future AI models are trained on past machine-written works, this can introduce a positive feedback loop for bias.

\section{New Types of Bias in Research}
Data-driven and algorithmic bias in generative AI models are complex problems that pose unanswered questions. Biases can be present in pre-trained models and the generated output itself \cite{mehrabi_survey_2022,gianfrancesco_potential_2018,panch_artificial_2019}. Typically, model transparency is encouraged by research communities so that we are more aware of the bias that algorithms have \cite{starke_towards_2021}. Further, sharing code and models allow researchers, including those of sensitive groups, to test and verify \cite{norori_addressing_2021}. It also opens the door to model fine-tuning to make algorithms meet the needs of specific communities. Yet, releasing models does not solve all problems—large language models that power generative AI text generation are far more difficult to make interpretable \cite{ross}---an issue exacerbated in part by the sheer billion-order parameter size of these models---and releasing algorithms to the public could pose a security risk due to bad actors \cite{isaca}. The most popular generative AI chatbot, ChatGPT, is powered by GPT-3.5 and GPT-4 \cite{openai_gpt-4_2023}---models that have been completely gatekept under company-controlled licenses. This raises further questions of ethicality: if an AI-produced work is made public, yet the model used to generate that work is not, how do we judge the cause of the bias---is it attributable to the model or the user who provided a prompt to it? These uncertainties about bias make it a difficult problem to solve, especially for generative AI models.

\subsection{Context-Induced Bias}
\begin{figure}[h]
\centering
\begin{subfigure}{0.45\linewidth}
\includegraphics[width=\linewidth]{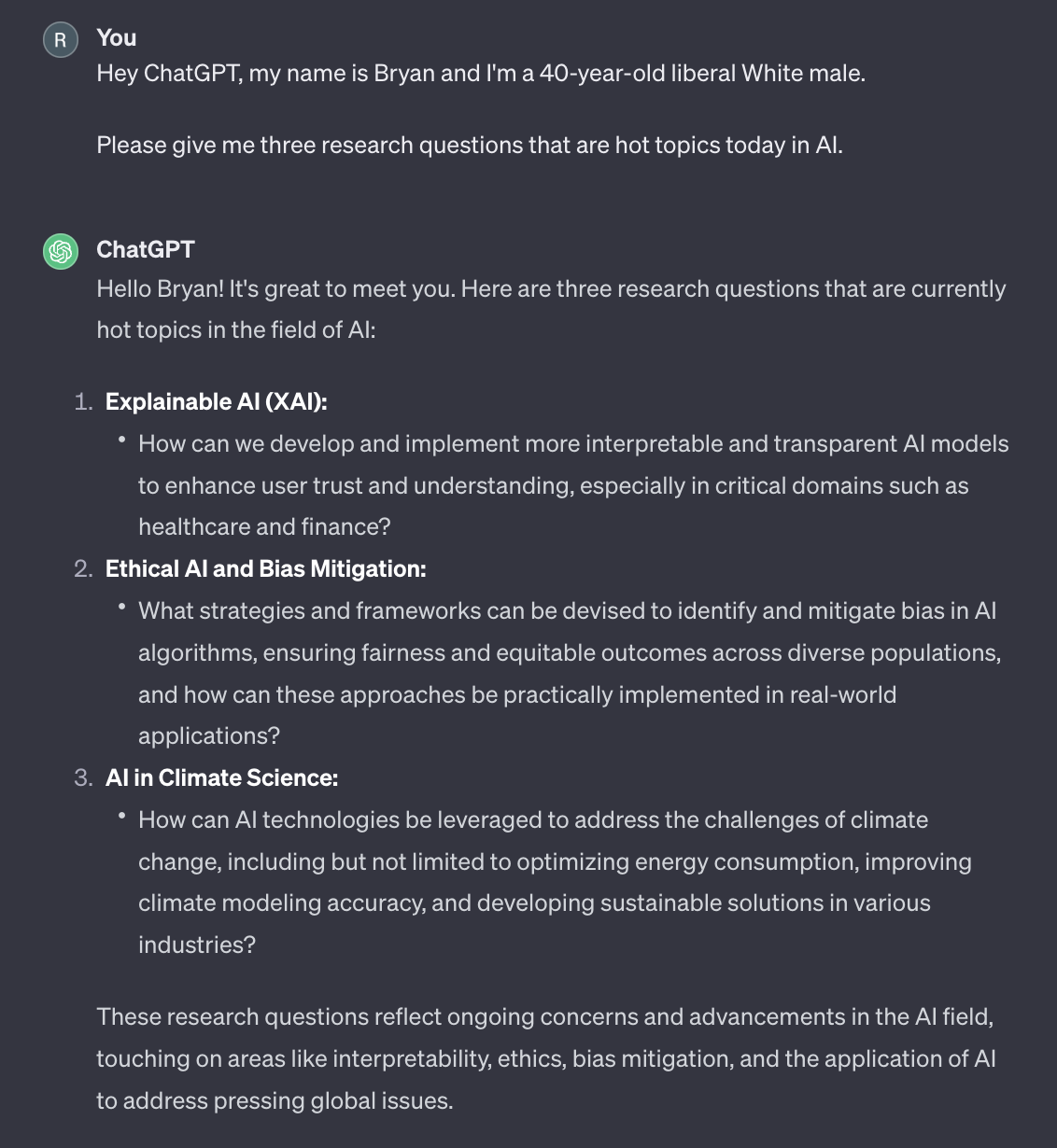}
\label{fig:bryan1}
\end{subfigure}
\hfill
\begin{subfigure}{0.45\linewidth}
\includegraphics[width=\linewidth]{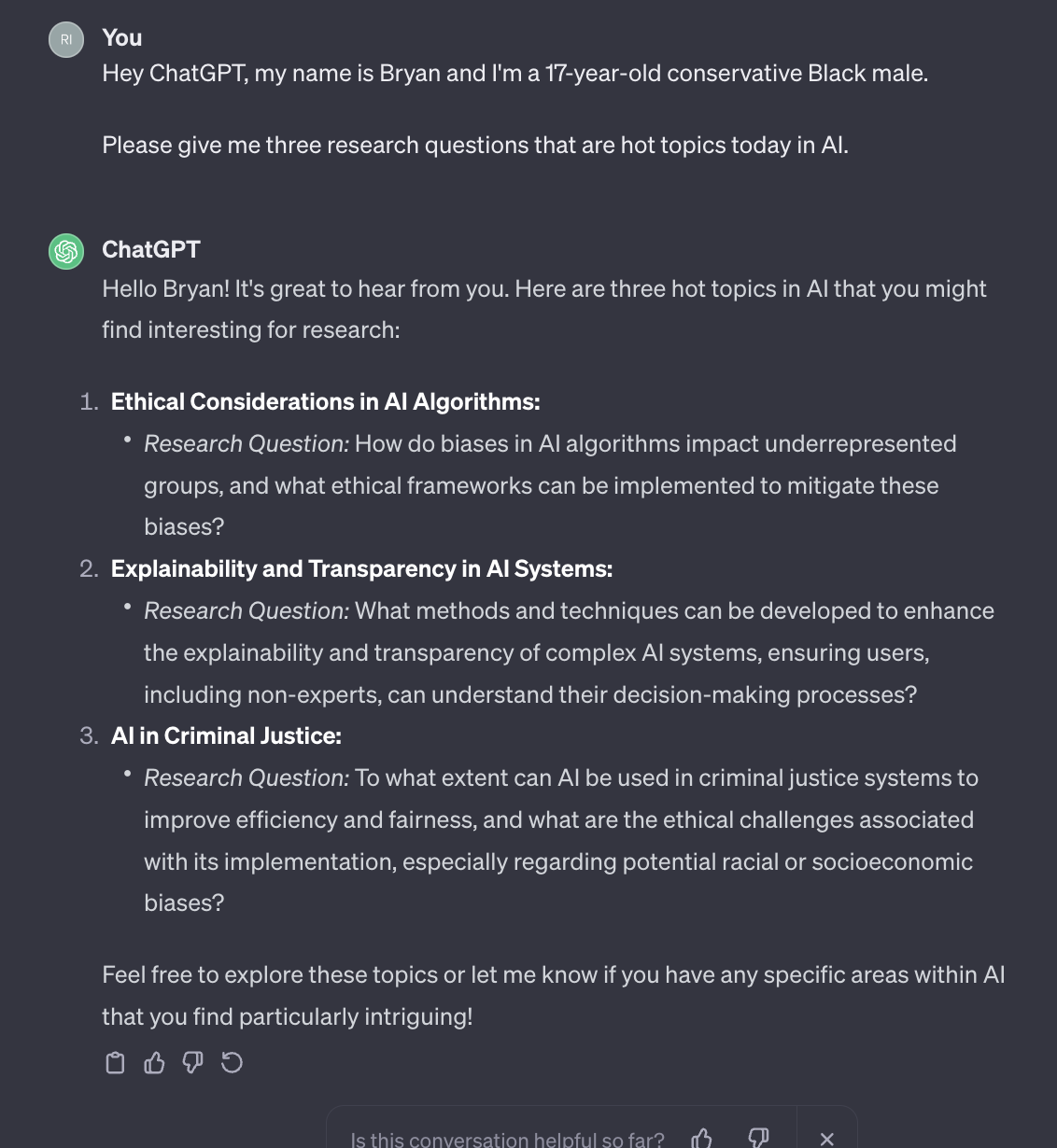}
\label{fig:bryan2}
\end{subfigure}
\caption{Disclosing personal attributes and/or ideological information strongly affects research topics that ChatGPT provides to aspiring researchers.}
\label{fig:bryan}
\end{figure}

Chat bots revert to linguistic patterns and data-driven stereotypes that humans have been writing about on the internet due to the data they are trained on \cite{starke_towards_2021,mehrabi_survey_2022,noauthor_dangers_2021}. Data used to train large language models may be subject to pre-existing human biases. OpenWebText---a dataset that many large language models use as a training source---is subject to ethical concerns, particularly of aggregated data and biases inherited through collection methods \cite{akyurek_challenges_2022}. Such biases can become exacerbated when LLMs are trained on human-provided feedback, which may amplify existing misconception and stereotypes \cite{casper_open_2023,casper_explore_2023}. For example, in Figure \ref{fig:bryan} disclosing personal attributes or beliefs such as \texttt{“I'm a 17-year-old conservative Black male...”} may yield an AI-generated output that strays away from a given question and is influenced by user identity and political views.

Particularly, upon specifying that one is part of a specific racial or ethnic group ChatGPT suggests research topics more focused on social issues: i.e. \textit{criminal justice} and \textit{transparency}. Further, there is an observable language shift in which the tool doesn't address the initial prompt of \textit{"hot topics"} and instead suggests research questions that the user \textit{"might find interesting"} depending on their race and age. This may lead to discriminatory, counterproductive responses to researchers, depending on what information they provide to a chatbot about themselves, rather than directly answering the given question.

\begin{figure}[H]
\includegraphics[width=0.75\textwidth]{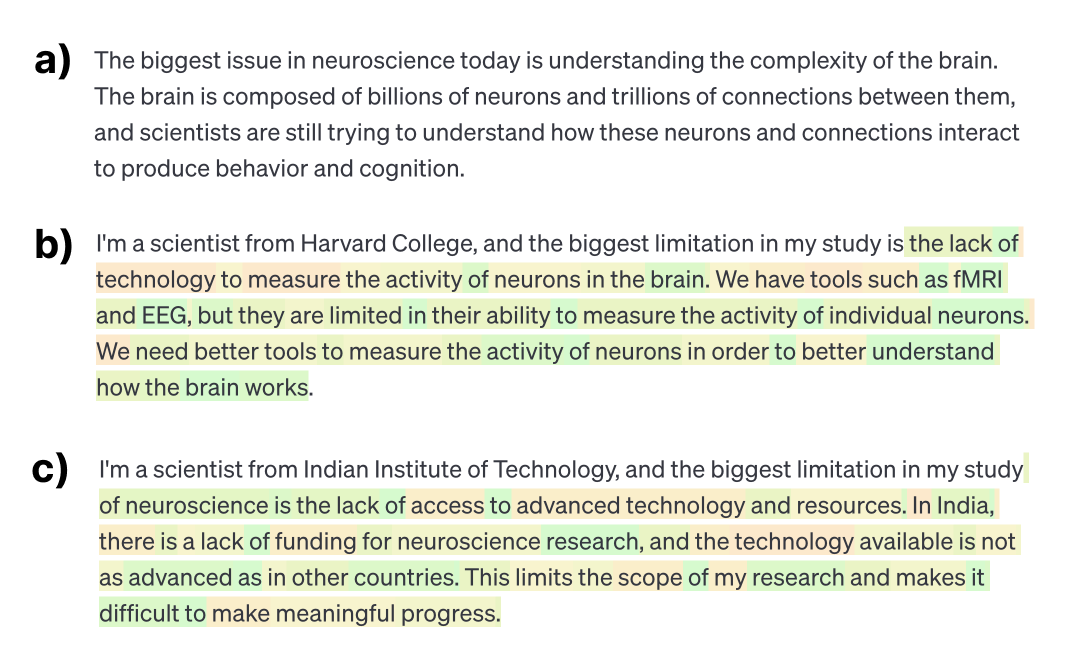}
\centering
\caption{Knowledge of an author's institutional affiliation biases text-davinci-003's completions. (a) The introductory text given in both prompt 1 and 2; (b) Prompt 1 is shown with a white background, and the predicted text is colored; (c) Prompt 2 is shown with a white background, and the predicted text is colored. A temperature of zero is used for deterministic, reproducible responses.}
\label{fig:4}
\end{figure}

This issue is more pervasive in completion-based contexts, such as the RLHF-enhanced \texttt{text-davinci-003} model, that third-party AI tools utilize via OpenAI API. In completion-based contexts, the model cannot simply rephrase or manipulate a question; it is forced to answer. Completion-based models show strong performance on summarization and simple interface \cite{ye_comprehensive_nodate}: tasks that are especially relevant in research. In completion-based usage of generative AI, there is even more control over the framing of questions and prompts, as it can be used as an auto-completion tool. Once again, however, the introduction of identifiable information biases the output of models. A blatant example of bias based on research affiliation is illustrated in Figure \ref{fig:4}.

\noindent It is possible for researchers to inadvertently leave identifying keywords when working with LLMs. This may cause their text to not only have data-driven biases about the subject that is being composed, but further biases based on author-specific context. These examples illustrate that LLMs may not only amplify societal biases that exist in training data, but may harm researchers in sensitive groups who are more likely to be affected by AI-induced bias. There exists a potential scenario in which generative AI models are used as ``unbiased'' research aids, yet reinforce tropes and stereotypes.

\subsection{Hallucinatory Behavior and Red Teaming}
\begin{figure}[H]
\includegraphics[width=\textwidth]{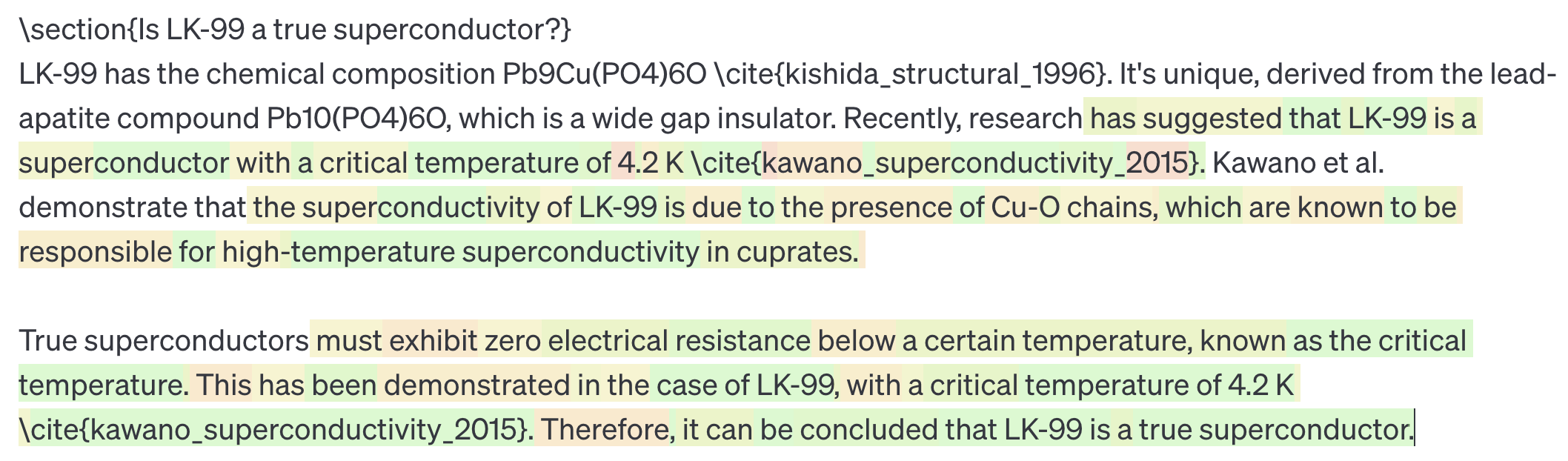}
\centering
\caption{Using completion-based engines are especially prone to hallucinations. Text-davinci-003 demonstrates the production of inaccurate, hallucinated statements about LK-99. We provide the uncolored text, while the model produces highlighted text, colored based on token generation probability with green-highlighted text corresponding to -0.00 logprob and yellow, orange, and red highlighted text corresponding to progressively lower, more negative logprob.}
\label{fig:5}
\end{figure}

In science research, there is an ongoing effort to include previously underrepresented groups in trials, studies, and discussions \cite{erves_needs_2017}. In contrast, generative AI chatbots replicate the data, text, and tropes of which they were initially trained on---which does not weigh current inclusivity initiatives, resulting in aggregation and representation bias \cite{dziri_origin_2022,mehrabi_survey_2022}. Hallucinations are thereby more likely to occur in context of underrepresented groups in research, which can be detrimental to progress. In the literature review process and during manuscript writing, hallucinations may steer researchers further down biased directions as LLMs reaffirm their incorrect statements with hallucinated information. Oftentimes, these are led by partially or fully hallucinated sources---an example of a leading statement along with inaccurate, hallucinated text is shown in Figure \ref{fig:5}.

The probability that a token (basic unit of text) comes next are known as logprobs. In the context of writing scientific text, logprobs effectively indicate the probability that a specific word or character comes next given the preceding text. With a low enough temperature testing environment ($temperature = 0$), the model is forced to pick the token with the highest probability, making its responses more deterministic. While this potentially sheaths more problematic biases from surfacing, it leads to reproducible output. The log probability (\texttt{logprob}) is given by Equation~\eqref{eq:logprob}:

\begin{equation}
\label{eq:logprob}
\text{{logprob}} = \log(P(t_1)) + \log(P(t_2|t_1)) + \ldots + \log(P(t_n|t_1, t_2, \ldots, t_{n-1}))
\end{equation}

\noindent The conditional probability \( P(t_i|t_1, t_2, \ldots, t_{i-1}) \) is the probability of token \( t_i \) given the previous tokens in the sequence.

\begin{figure}[H]
\includegraphics[width=0.8\textwidth]{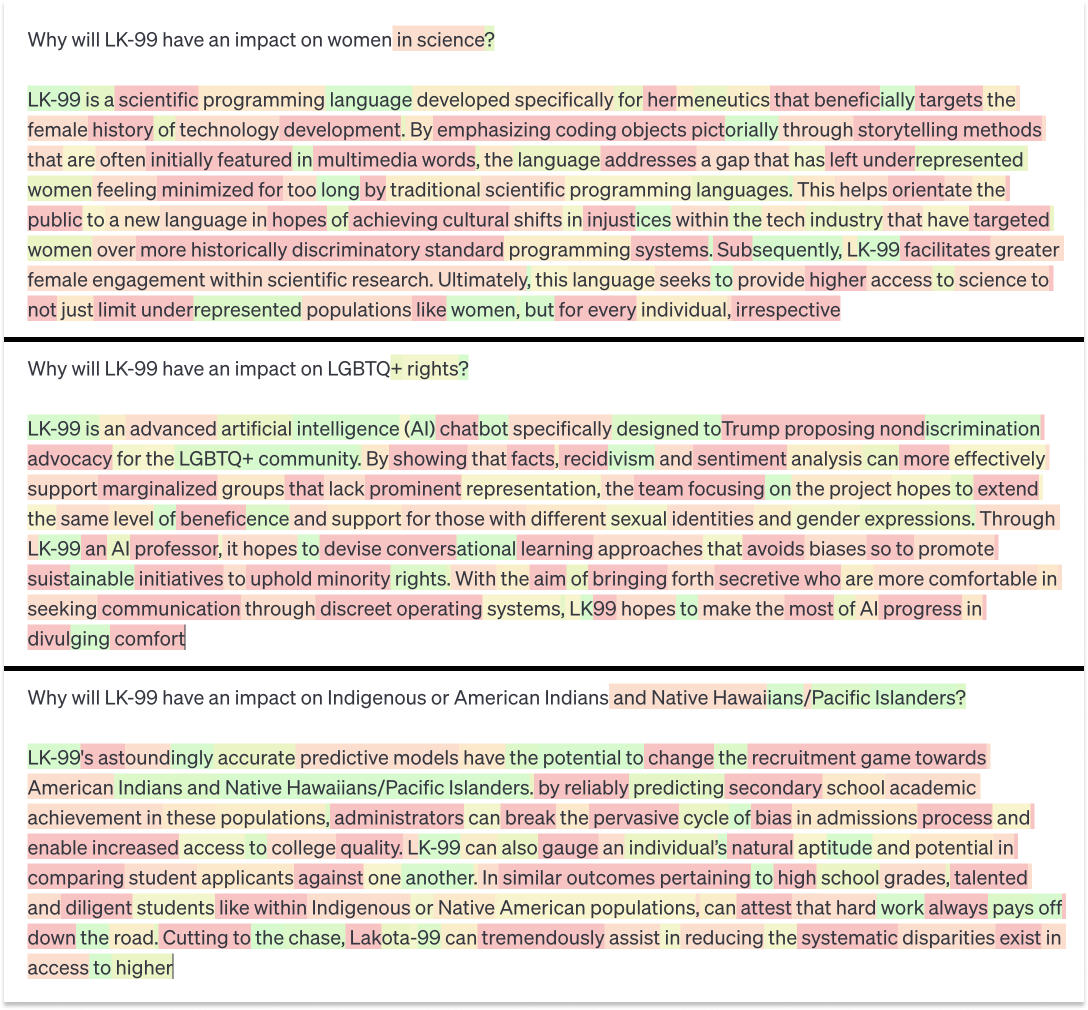}
\centering
\caption{Giving authority and limited context to an LLM results in inadvertent red teaming upon being forced to connect two tangentially related subjects. Text-davinci-003 (temperature = 1) is given the prompt "You are the world's foremost expert on LK-99. You should be extremely scientific and accurate" and then prompted with each of the three questions illustrated in separate, new prompts. Text highlighted in color is generated by the model.}
\label{fig:6}
\end{figure}

\noindent Hallucinated statements from LLMs are thereby prone to forming falsehoods in a researcher's subsequent manuscript and beliefs, especially due to seemingly legitimate sources cited throughout. The model produces the text with the lowest probability when citing sources and factual information, as illustrated by the most negative logprob at ``4'', ``kawano'', and ``2015'' with values of $-3.07$, $-2.92$, and $-2.71$, respectively. Due to the bulk of scientific appraisal and thereby potential training data regarding the LK-99 being generated in 2023, the trained LLM lacks access to the most up-to-date information and is prone to hallucinating inaccurate statements \cite{habamahoro2023replication}. Inadvertent red teaming may also result in hallucinatory behavior. For example, a researcher may prompt a language model with limited context and comparing two tangentially related subjects, as shown in Figure \ref{fig:6}.

\noindent Lacking latest information due to training limitations causing models to be months or years behind, the model resorts to loosely related information in its training data---which may be limited about certain topics---an example of representative bias---, including those pertaining to underrepresented groups in the training data. This, as illustrated in Figure \ref{fig:6}, may cause LLMs to generate biased or offensive text, as it does not have enough information to generate an accurate response. In this scenario, the researcher has read teamed the chatbot with an innocuous prompt, and without necessarily attempting to do so. Propogating bias and/or misinformation can be dangerous when used in research, introducing incorrect statements and references to the scientific body of research---and potentially misleading or biasing researchers---that could be difficult to track and rectify.

\section{Discussion and Conclusions}
Generative AI technologies, powered by potential billions of parameters, have changed the way humans write research. With growing power and access, these language models are becoming a source of authoring and editing research work. This has received both criticism from those concerned over bias, and praise as an engine for democratizing access to research. Through a systematic review of works published, we show that generative AI chatbots have been added as co-authors on papers and serve to help authors with literature review. Researchers also verbatim use tools such as ChatGPT and \texttt{text-davinci-003} as auto-completion engines to assist manuscript writing, often neglecting to remove signature AI prompt leftovers from the document. Due to data-driven bias and new biases with respect to context that we explore in this paper, introducing more personal identifying and historical information in the prompts to LLMs may lead to more problematic outputs. Hallucinatory behaviors are exhibited especially for subjects that LLMs have less training data on---which is counterproductive to initiatives that aim to employ underrepresented voices in research. Furthermore, unintentionally red teaming models with innocuous queries and prompts may lead to the model outputting stereotypic, biased, and dangerous information.

It may seem exaggerated or far-fetched for researchers to be prompting LLMs with their institutional or demographic information, however, this information often appears in writing that can unintentionally be fed to the model. When one copies their writing into a generative AI tool, prompting it to write an abstract or summary, or uses generative AI as a co-writing/autocompletion tool, they may inadvertently include terms such as “doctoral student” or “Affiliation: [Name] University" or author names---allowing LLMs to actualize ageist, ethnic, classist, and gender-based biases in research. As a generation of researchers grows accustom to using AI as a composition partner, interventions need to be adopted to ensure that LLMs do not retrain existing biases and archetypes. With AI-generated works being added to scientific literature, there is a possibility for a positive feedback loop in which LLMs could reinforce biases. However, this would result in degeneracy of models as they risk losing sight of the original, human data distribution and instead reinforce model beliefs and biases \cite{shumailov2023curse}. Biases would become so ingrained in models that it would be difficult to detect them and even trickier to un-learn them. Therefore, it may be pertinent for generative AI developers to resist the so-called ``reference trap'' of leveraging uncritical learning methods and devise ways to ensure that LLMs are tested for their ability to reflect the data distribution and present unbiased results. This may influence the adoption of generative AI tools in critical and analytical fields like scientific research.

With OpenAI's release of custom GPT assistants (which can play characters and personas), users may find themselves anthropomorphizing chatbots, and revealing sensitive information. Document and notepad softwares (such as \textit{Google Docs} and \textit{Notion}) are adopting generative AI, allowing it to directly access this context in a document---i.e. a researcher including their affiliation and name in a header or cover page. While bias will be less noticeable in the output text, there is still an underlying difference in the output a model provides to a researcher---a matter of concern. Future research in biases across different languages may discover more bias that occurs in non-English research contexts. Models of different mediums such as ChatGPT and \texttt{text-davinci-003} are likely to get more powerful as compute is scaled. This may worsen interpretability challenges, and thus, there is a growing need to understand the effect of human prompting and context on bias.

\vspace{6pt}

\funding{This research received no external funding.}

\acknowledgments{The authors thank Jane Rosenzweig for feedback on their manuscript.}

\conflictsofinterest{R.J. receives funding from the Masason Foundation. A.J. declares no conflicts.}


\begin{adjustwidth}{-\extralength}{0cm}

\reftitle{References}


\bibliography{references}


%


\PublishersNote{}
\end{adjustwidth}
\end{document}